\documentclass[conference]{IEEEtran}
\IEEEoverridecommandlockouts
\usepackage{cite}
\usepackage{amsmath,amssymb,amsfonts}
\usepackage{algorithmic}
\usepackage{graphicx}
\usepackage{textcomp}

\usepackage{CJKutf8}
\usepackage{xcolor}

\usepackage{multirow}
\usepackage{multicol}

\usepackage{graphicx}
\usepackage{subfigure} 
\usepackage{float}

\usepackage{booktabs}
\usepackage{amsmath}
\usepackage{setspace}
\usepackage{array,caption}
\usepackage[labelfont=bf]{caption}
\usepackage{threeparttable}
\usepackage{tabularx}
\usepackage{hhline}
\usepackage{booktabs}

\usepackage{longtable}

\usepackage{makecell}

\usepackage{hyperref}

\usepackage{pifont}

\usepackage{ragged2e}
\def\BibTeX{{\rm B\kern-.05em{\sc i\kern-.025em b}\kern-.08em
    T\kern-.1667em\lower.7ex\hbox{E}\kern-.125emX}}

\usepackage{float}
\usepackage{booktabs}
\usepackage{multirow}

\begin{document}
\begin{CJK}{UTF8}{gbsn}

\title{A Systematical Evaluation for Next-Basket Recommendation Algorithms
}

\author{
\IEEEauthorblockN{
    Zhufeng Shao\IEEEauthorrefmark{1},
    Shoujin Wang\IEEEauthorrefmark{2},
    Qian Zhang\IEEEauthorrefmark{1},
    Wenpeng Lu\thanks{ \ding{41} Wenpeng Lu is the corresponding author.}\IEEEauthorrefmark{1}\textsuperscript{(\ding{41})},
    Zhao Li\IEEEauthorrefmark{3},
    Xueping Peng\IEEEauthorrefmark{4}
}
\IEEEauthorblockA{
    \IEEEauthorrefmark{1} School of Computer, Qilu University of Technology (Shandong Academy of Sciences), Jinan, China\\ 
    \IEEEauthorrefmark{2} The Data Science Institute, University of Technology Sydney, Sydney, Australia \\
    \IEEEauthorrefmark{3} Shandong Evay Info Technology Co., Ltd., Jinan, China\\Shandong Computer Science Center (National Supercomputer Center in Jinan), Jinan, China  \\
    \IEEEauthorrefmark{4} Australian Artificial Intelligence Institute, University of Technology Sydney, Sydney, Australia \\
    Email: zhufengshao7@gmail.com, shoujin.wang@uts.edu.au, qianzhang9706@gmail.com,\\ lwp@qlu.edu.cn, liz@sdas.org, xueping.peng@uts.edu.au}
}







\maketitle

\begin{abstract}

Next basket recommender systems (NBRs) aim to recommend a user's next (shopping) basket of items via modeling the user's preferences towards items based on the user's purchase history, usually a sequence of historical baskets. Due to its wide applicability in the real-world E-commerce industry, the studies NBR have attracted increasing attention in recent years. NBRs have been widely studied and much progress has been achieved in this area with a variety of NBR approaches having been proposed. However, an important issue is that there is a lack of a systematic and unified evaluation over the various NBR approaches. Different studies often evaluate NBR approaches on different datasets, under different experimental settings, making it hard to fairly and effectively compare the performance of different NBR approaches. To bridge this gap, in this work, we conduct a systematical empirical study in NBR area. Specifically, we review the representative work in NBR and analyze their cons and pros. Then, we run the selected NBR algorithms on the same datasets, under the same experimental setting and evaluate their performances using the same measurements. This provides a unified framework to fairly compare different NBR approaches. We hope this study can provide a valuable reference for the future research in this vibrant area.


\end{abstract}

\section{Introduction}
Recent years have witnessed the great success of Recommender Systems (RSs) in many different real-world applications, such as E-commerce, stream media and online retail industry \cite{hu2020modeling, qin2021world,lu2022aspect,qian2022,lu2019graph,lu2021sentence}. RSs have become a fundamental tool for users to make right choices from massive and redundant contents, products and services in an effective and efficient way \cite{wang2021survey}. As one of the most commonly used and practical RSs, next-basket recommender systems (NBRs), as a sub-area of RSs, have attracted increasing attention in recent years. 


Why next-basket recommender systems?
(1). Next-basket recommender systems (NBRs) are one of the most applicable RSs in the real-world shopping scenarios. The reason is that users often purchase a (shopping) basket of items rather than one single item in a shopping visit. For instance, Bob usually purchases a basket of daily products in his weekly shopping event. Therefore, NBRs can naturally match the real-world shopping scenarios since they aim to recommend a basket of items which are carefully selected to match a user's current demand and preference. 
(2). Although a variety of approaches for next-item recommendation task have been developed\cite{song2021next,wang2020modelling,wang2019modeling,wang2018attention}, they aim to recommend the next item within the current basket only via modeling the intra-basket correlations over items. As a result, they cannot be employed for the next-basket recommendation task in which a basket of inter-correlated items are recommended via modeling the sequential dependencies over a sequence of baskets. Therefore, new theories and approaches are in demand for next-basket recommendation, making NBR a significant research topic.
(3). NBRs predict the next basket of items which mostly interest a user by capturing the user's preference from her/his purchase history, namely a sequence of baskets purchased recently. As a result, both the user's long-term and short-term preferences can be well modeled for more accurate recommendation.



Given a user's historical transaction records, usually a sequence of shopping baskets, a NBR aims at predicting the next basket of items that a user would like to purchase by modeling the sequential dependencies over the sequence of baskets. A variety of existing studies on next-basket recommendation have emerged with great success. For example, a Markov Chain (MC) based approach (Rendle et al.) \cite{r2} has been proposed to capture low-order dependencies over baskets for next-basket recommendation, embedding-based approaches (Wan et al.; Wang et al.) \cite{r5,r15} have been developed to use distributed representation to predict the next basket, recurrent neural network (RNN) based approaches (Yu et al.; Hu et al.; Le et al.; Bai et al.; Qin et al.) \cite{r1, r12, r13, r16, qin2021world} were proposed to capture higher-order dependencies across baskets. In addition, intention-driven approaches (Wang et al.) \cite{wang2020intention, wang2021intention2basket} have been proposed to model the heterogeneous intentions contained in the historical sequences of purchased baskets to recommend next basket that satisfies user's different intentions.

In another line of work, a K-nearest neighbor (KNN) based approach (Hu et al.) \cite{hu2020modeling} was proposed to exploit personalized item frequency information for improving the performance of NBRs. Similarly, another approach based on KNN and collaborative filtering (CF)  (Faggioli et al.) \cite{r14} was proposed to model the recency of items for NBRs.


Although there are many studies on NBR existing in the literature and most of them have achieved great success, one critical issue has attracted much attention: there is a lack of a systematic and unified framework to comprehensively and systematically categorize NBR approaches and evaluate them in a rigorous way. As a result, it is not very clear what is the latest research progress in this vibrant area and how the different NBR approaches really perform. There is also a lack of problem formalization and unified experimental settings for NBR research, making it hard to compare different approaches in a fair way\cite{wang2022trustworthy}. Also, the inconsistency on the selection of datasets, baselines commonly exist in various studies.

The aforementioned gaps in the existing NBR research motivate us to conduct a systematic investigation on various NBR studies in the literature in both a quality and a quantity way. To be specific, in this work, we first provide a formal problem statement for the NBR problem, then summarize the research progress in NBR area by categorizing and comparing the representative and state-of-the-art NBR approaches. Afterwards, we set up a unified experimental environment and conduct a systematic empirical study on most representative NBR approaches which are carefully selected. 

To the best of our knowledge, this is the first work to systematically investigate the NBR approaches in the literature. The main contributions of this work are summarized below: 
\begin{itemize}
\item We provide a novel taxonomy to well categorize and organize a variety of representative NBR approaches. As a result, an overview of the current progress of research in NBR area has been provided.     

\item We conduct a comprehensive and systematic empirical study on the representative and state-of-the-art NBR approaches. This provide a unified evaluation for various NBR approaches to well compare their performance in a rigorous and fair way. We hope this work shade some light for the future studies in this vibrant area.


\end{itemize}

\section{RELATED WORK}


In this section, we first clarify the difference between next-basket recommendation and bundle recommendation which is a task quite relevant to but different from next-basket recommendation. Then, we review some existing surveys and reviews related to the topic of next-basket recommendation.

\subsection{Next Basket Recommendation vs. Bundle Recommendation}

There have been some studies on other tasks relevant to next-basket recommendation, among which, bundle recommendation \cite{r24} is the most typical one. Although superficially similar, next-basket recommendation and bundle recommendation vary in settings and assumptions while researchers often mix up them. Therefore, it is necessary to conduct a comparison between them and thus differentiate next-basket recommendation from bundle recommendation.

Next basket recommendation and bundle recommendation are built on basket data and bundle data respectively. So it is necessary to first clarify the difference between basket data and bundle data. A basket is a list of items purchased by a user in one shopping visit. There is usually no clear order over the items within a basket. The items in one basket may be correlated or uncorrelated, which depends on the specific cases. Given a sequence of baskets, they are often sorted in ascending order in terms of a shopping visit.
In comparison, a bundle is the integration of two or more highly correlated items \cite{r25}. The items in a bundle are unordered and are often similar or complementary with each other \cite{hao2020p}. When regarding bundles as a sequence, bundle recommendation usually gives the order by sorting the bundles in terms of their prices in a descendant order.

With the user's transaction history, namely a sequence of historical baskets, next basket recommendation is often formalized as a sequential prediction task, i.e., next basket recommendation targets to predict what items the user will buy in her/his next shopping visit. In such a case, the sequential dependence over a user's historical purchased baskets was fully modelled and employed. For example, when we know a user has purchased a printer in her current shopping basket, we can recommend some relevant items such as printing paper and toners as her next basket. In comparison, bundle recommendation\cite{chen2019matching} is formalized as an optimization problem which aims to select a set of optimal and correlated items from the massive candidate items to accomplish a certain consumption goal of a user. Different from next-basket recommendation, no obvious sequential dependency will be considered and modelled in bundle recommendation. For instance, given the high relevance between iPhone and AirPods, we can recommend them together as a bundle to a user who likes Apple products without modeling any sequential purchase behaviors.




\subsection{Related Surveys}

There are also many other studies related to next-basket recommendation, e.g., session-based recommendation and sequential recommendation. For session based recommendation, Wang et al.\cite{wang2021survey,wang2020hierarchical} provided a comprehensive and systematic survey to formally define the research problems, illustrate the main research challenges, review the main research progress and point out the promising research directions in this area. Ludewig et al. \cite{r26} provided a systematic empirical study on a variety of session-based recommendation algorithms to comprehensively compare their recommendation performance. For sequential recommendation, Wang et al. \cite{r7} summarized the key challenges, progress and future directions in this important research area. Fang et al.\cite{r9} summarized and compared deep learning based approaches for sequential recommendation. 

However, although relevant, session-based recommendation and sequential recommendation are totally different from next-basket recommendation. The reason is that session-based recommendation and sequential recommendation are based on session data and sequence data respectively, in which a set of items (a session) or a sequence of items are taken as the input of the recommendation algorithm. As a result, there is no clear basket structure inside session data or sequence data. In another words, they mainly recommend the next item within the same basket by modeling the intra-basket dependencies. In contrast, the input data for next-basket recommendation is a sequence of baskets, and an NBR aims to recommend the next basket via modeling the inter-basket dependencies.           

Although a series of review work and empirical studies have been done in the areas of session-based recommendation and sequential recommendation\cite{guo2021sequential,hu2017diversifying,wang2022exploiting}. There is no systematic work to provide a comprehensive investigation on the various studies in the NBR area and there is not a unified and rigorous evaluation on the different NBR algorithms. Given the increasing popularity and potential of NBR and the emerging research progress in this area, a systematic and thorough summarization, evaluation and analysis of NBR approaches is in urgent demand. As the first attempt, this paper explores the field of NBR with an emphasis on the problem statement, classes of existing NBR approaches, benchmarking evaluation and provides further insights of future prospects in the area.



\section{PROBLEM STATEMENT}

In this section, we formally define the research problem of NBR and then discuss the main work mechanism of NBR. 



\begin{figure}[b]
	\begin{centering}
		\includegraphics[width=0.47\textwidth]{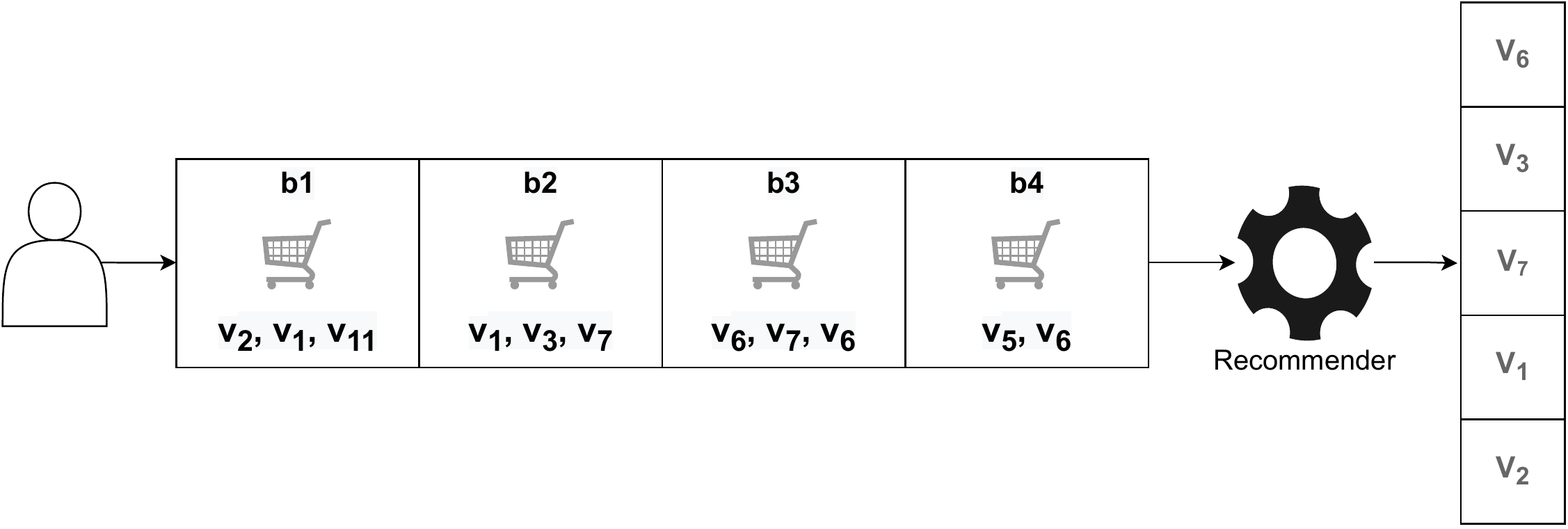}
		\caption{A running example for next basket recommendation.}
		\label{fig:NBRfigure}
	\end{centering}
\end{figure}

There are different definitions on NBR task based on different domains in the collected papers. For example, a basket could be defined not only as a series of products purchased in one transaction event in E-commerce domain, but also as a set of places visited in a trip\cite{r13,r19} in the tourism domain. To simplify the concepts, we call both a product and a place as an item in this paper. Consequently, a \textit{basket} is defined as a collection of items which have been purchased together in one transaction event by a given user. In the next-basket recommendation task, often given a sequence of historical baskets purchased by a user recently as the input, a next-basket recommender system is built and trained on such input data to predict the next basket of items which mostly interest the user via modeling the inter-basket dependencies. Usually, the prediction is performed in the form of generating a personalized ranked list of items (as depicted in Figure \ref{fig:NBRfigure}), which are supposed to form the next basket for the given user.



Now we follow the problem statement in \cite{wang2020intention} to formally define the research problem of next-basket recommendation. Given a transaction dataset $D=\{s_{1},...,s_{|D|}\}$ where $|D|$ is the total number of sequences, it contains a set of sequences of shopping baskets (called baskets for short). In $D$, each sequence $s=\{b_{1}, ..., b_{|s|}\} (s \in D)$ consists of a list of historical baskets purchased by a certain user and they are sorted in the order of purchase time. In each sequence, a basket $b=\{v_{1}, ..., v_{|b|}\} (b \in s)$ contains a collection of items which were purchased in one transaction event. All the items occurring in the whole dataset constitute the universal item set $V=\{v_{1},...,v_{|V|}\}$ while all the users occurring in the dataset form the universal user set $U=\{u_{1},...,u_{|U|}\}$.

Given a sequence of baskets $s=\{b_{1}, ..., b_t\}$, we pick up one basket, usually the last one $b_t$ as the target basket to be predicted, while all the other baskets occurring prior to $b_t$ will be taken as the corresponding context, denoted as $C_t = \{b_{1}, ..., b_{t-1}\}$. Note that in the prediction and recommendation stage, the target basket is unknown and needs to be predicted by the next-basket recommender systems. In a next-basket recommendation task, given a context $C_t$ of a user $u$, namely the purchase history of $u$, a next-basket recommender system aims to predict the corresponding user's choices for the $t^{th}$ basket, namely to generate a list of items which are most probably to appear in the user's next basket $b_{t}$. This can be formally defined below: 

\begin{equation}
b_t =  f(C_t, u).
\end{equation}

\begin{figure*} 
    \vspace{-20pt}
    \centering
        \subfigure[Distribution of publishing year]{
        \includegraphics[width=0.38\linewidth]{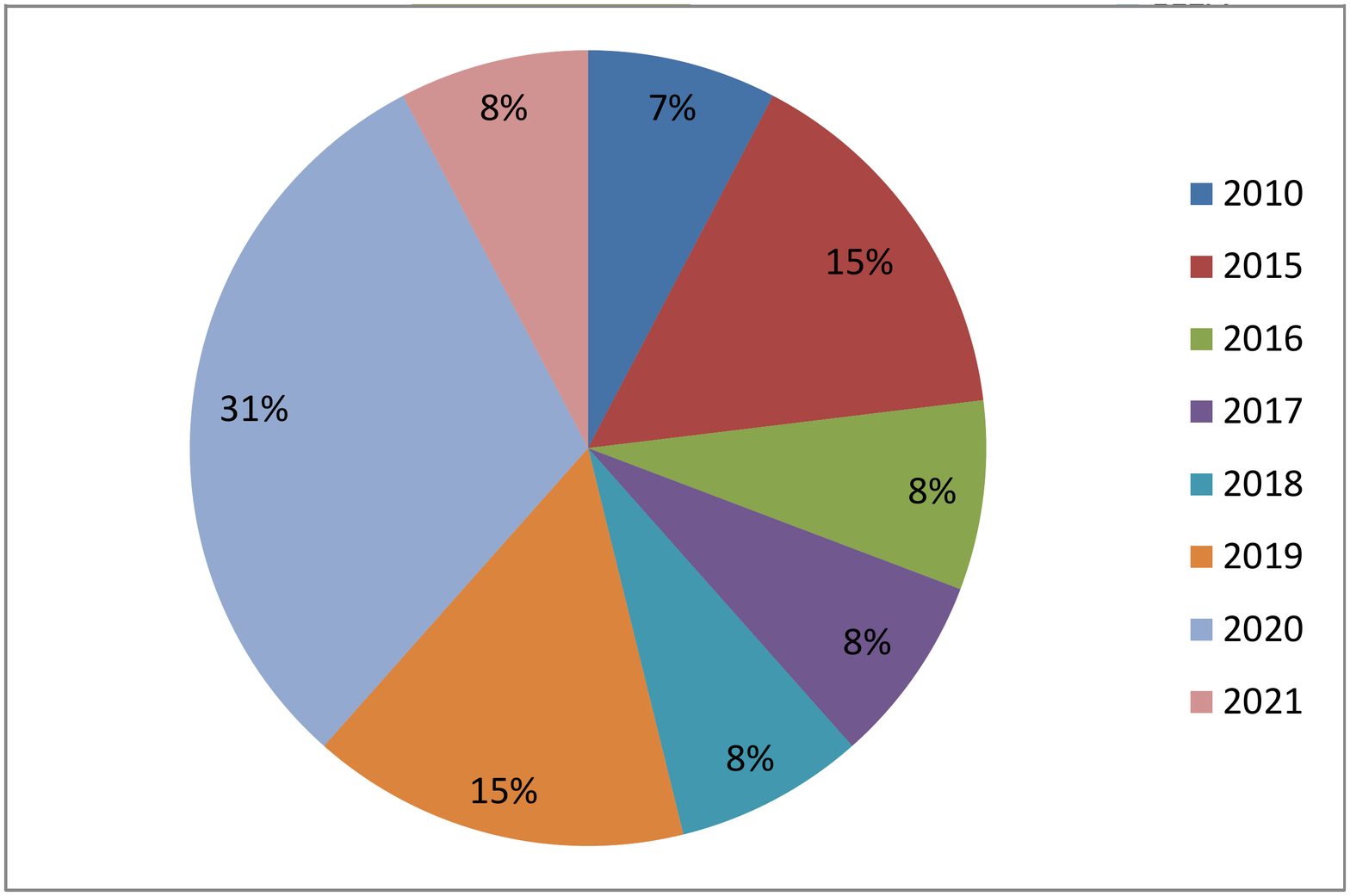}
        }
        \subfigure[Distribution of used experimental datasets]{
        \includegraphics[width=0.38\linewidth]{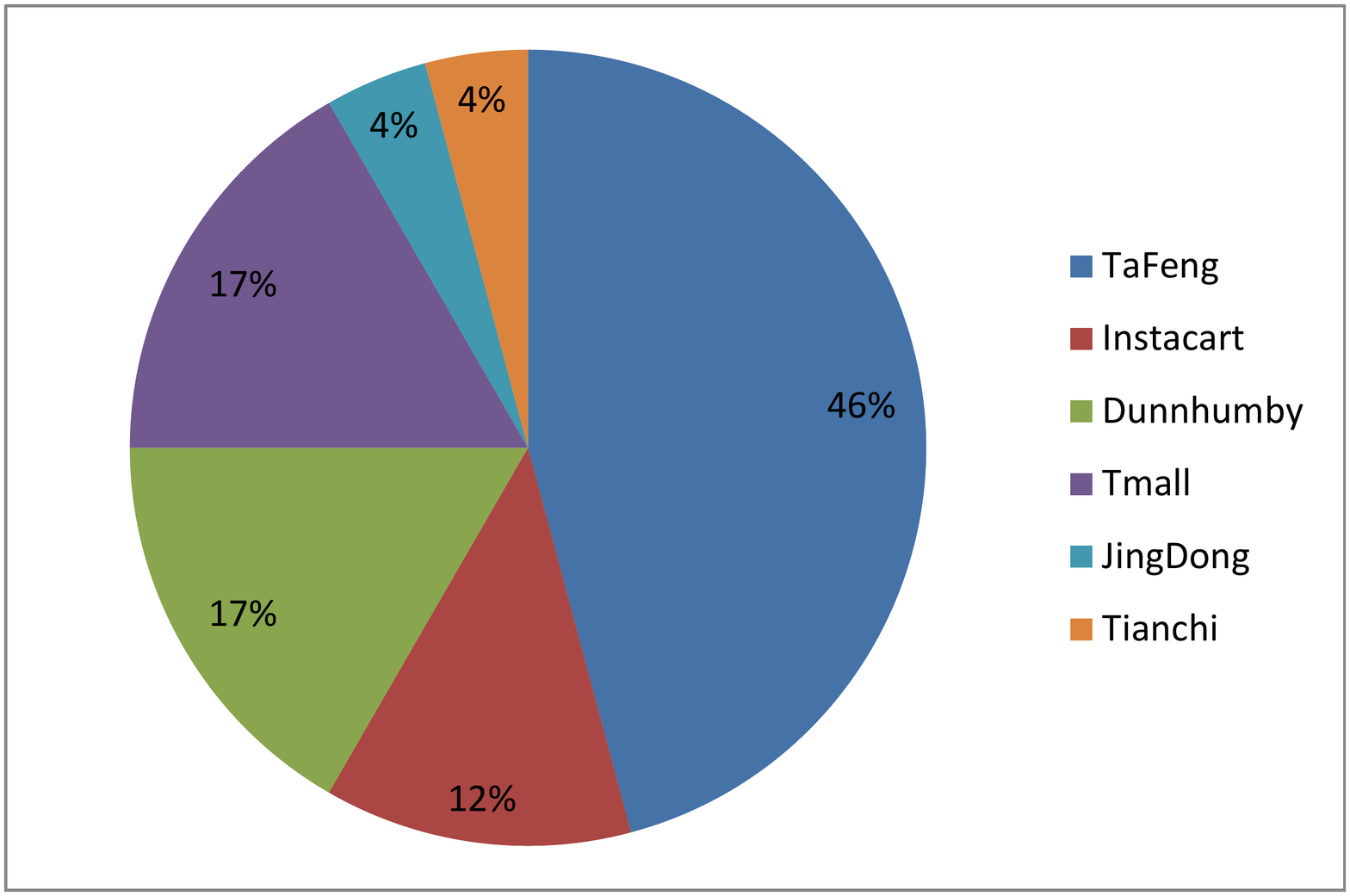}
        }
        \subfigure[Distribution of compared approaches]{
        \includegraphics[width=0.38\linewidth]{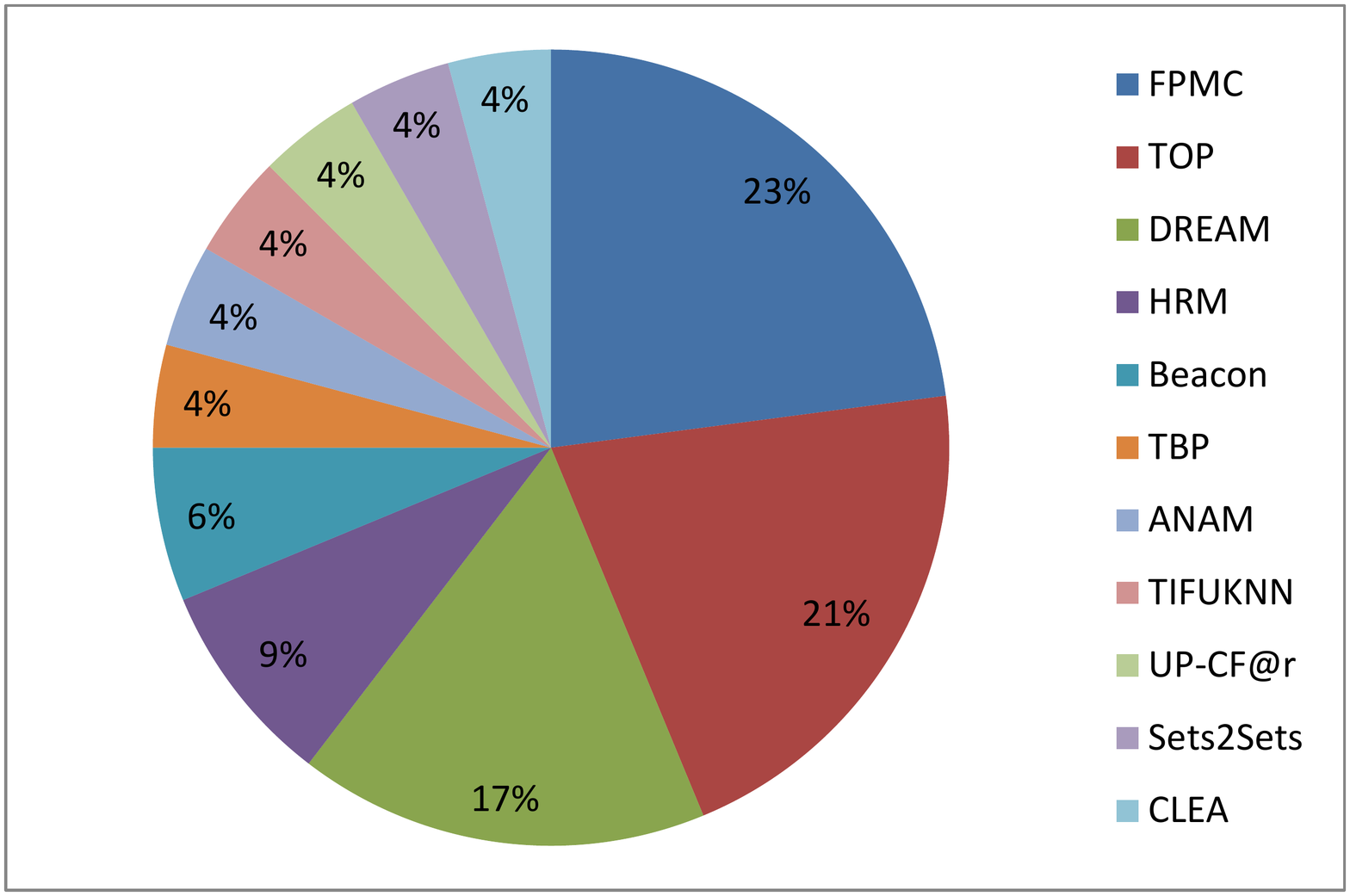}
        }
        \subfigure[Distribution of used evaluation metrics]{
        \includegraphics[width=0.38\linewidth]{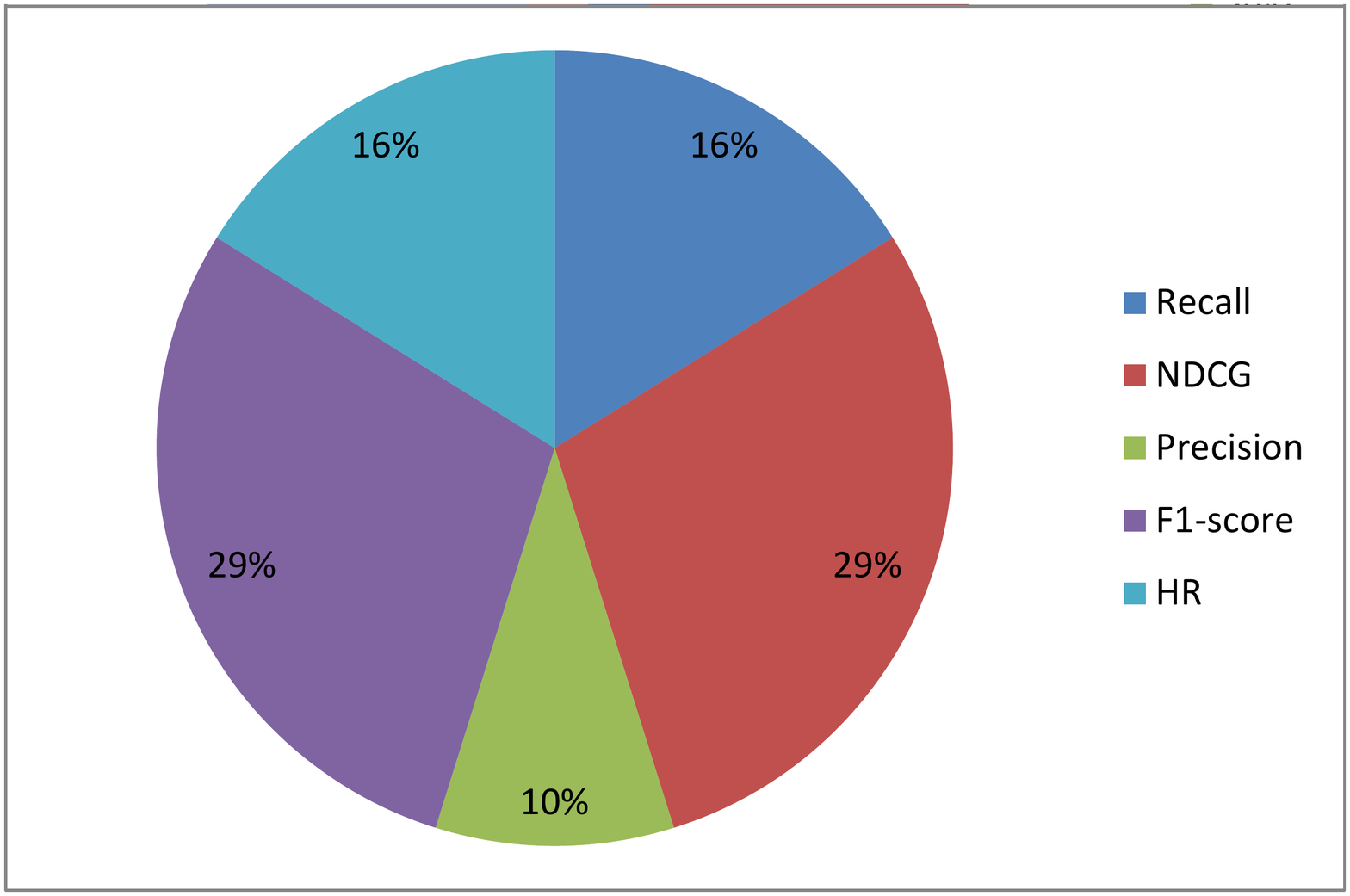}
        }
    \caption{(a) shows the distribution of publishing year of collected papers ;  (b) depicts the popularity of top-6 datasets; (c) shows the popularity of 11 compared NBR approaches; (d) shows the popularity of top-5 evaluation metrics.}
	\label{fig: summary_four_child} 
\end{figure*}

\section{Classes and Comparison of NBR Approaches}
\begin{figure}
    \begin{centering}
    \includegraphics[width=0.48\textwidth]{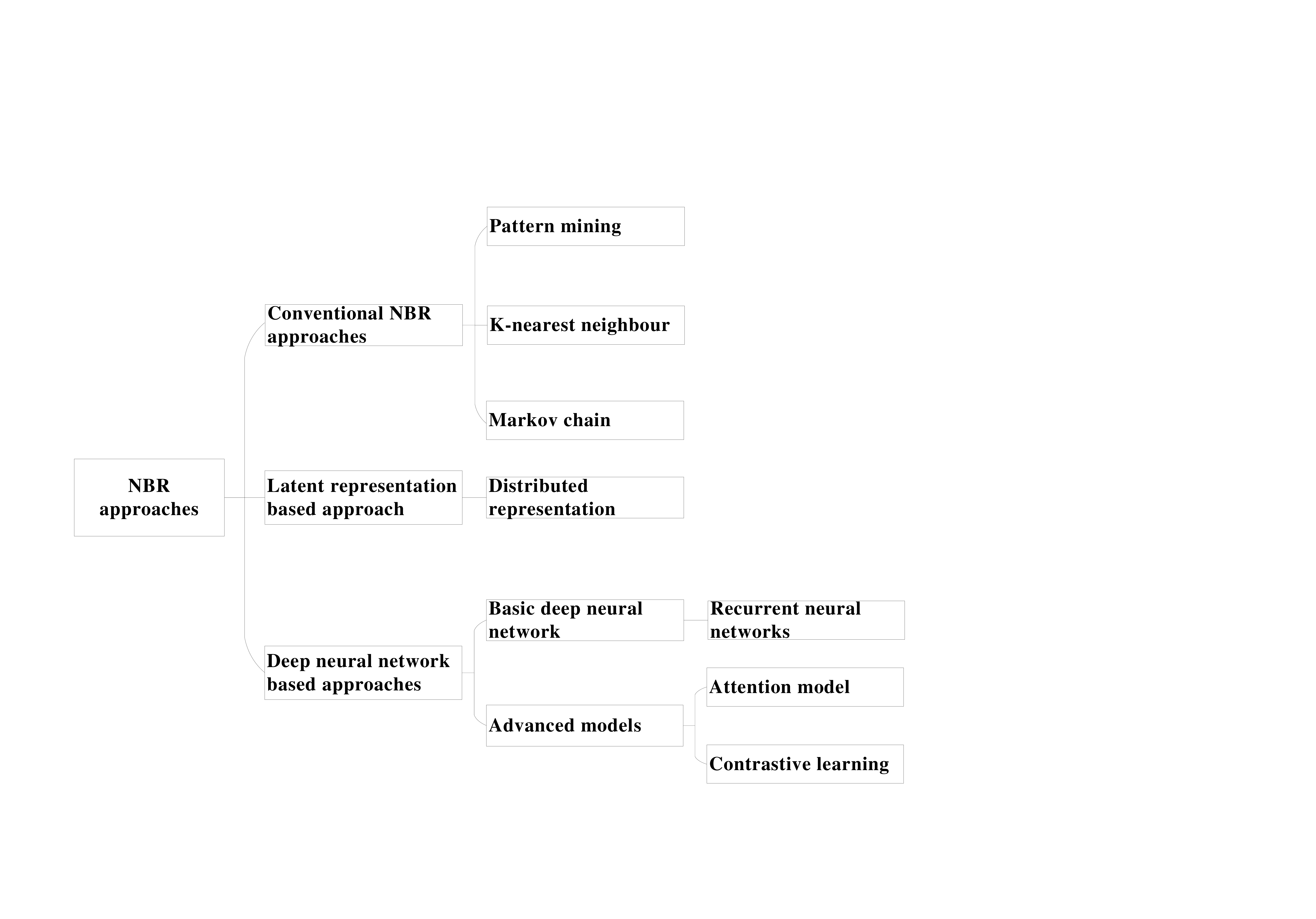}
    \caption{The classes of NBR approaches.}
    \label{fig:categorizationofNBR}
    \end{centering}
\vspace{-10pt}
\end{figure}

In this section, to provide an overview of the NBR research progress, we first propose a taxonomy to well classify the representative and state-of-the-art NBR approaches, and then compare the different classes of approaches systematically. Specifically, there are a variety of studies in NBR area and it is impossible to analyze each of them in detail. Therefore, in this section, we carefully select 13 most representative and highly cited works to analyze. To be specific, we first search the papers whose title contain the keyword \textit{next basket recommendation} in Google Scholar and then select those papers with equal or more than 10 citations. To provide a more straightforward view on the research progress in NBR, we conduct a bibliometric analysis over the selected literature. The distributions of publication year, experimental datsets, compared approaches and used evaluation metrics of the literature have been shown in Figure \ref{fig: summary_four_child}.





\subsection{A Categorization of NBR Approaches.}

As depicted in Figure \ref{fig:categorizationofNBR}, according to the utilized data mining or machine learning models and techniques, three major classes for NBR approaches are identified from the literature, i.e., (1) conventional NBR approaches which are built on conventional data mining approaches such as pattern mining and K-nearest neighbour (KNN); (2) latent representation approaches which are mainly built on representation learning techniques such as latent embedding; and (3) deep neural network approaches which are typically built on deep learning models including RNN. These three classes can be further divided into eight sub-classes. Specifically, conventional NBR approaches contain four sub-classes: pattern mining approaches, K-nearest neighbour approaches, and Markov chain based approaches; latent representation method mainly contains one sub-class, i.e., distributed representation; and deep neural network approaches contain two sub-classes: basic deep neural networks and advanced models. In addition, the sub-class of basic deep neural network mainly contains recurrent neural network based NBR. The sub-class of advanced models contains two sub-subclass, i.e., attention models and contrastive learning. In particular, except the approaches based on a single algorithm/model, there are some hybrid approaches which combine more than one algorithms/models. For example, ANAM\cite{r16} combines recurrent neural network and attention model and CLEA\cite{qin2021world} combines a contrastive learning framework and a recurrent neural network, etc.


\subsection{A Comparison of Different Classes of NBR Approaches}

Generally, conventional NBR approaches are straightforward and not so complicated since they are mostly based on conventional models and algorithms like pattern mining \cite{r10}, KNN models \cite{hu2020modeling} and Markov chain models \cite{r2}. As a result, they are mostly more efficient and easy to implement. And they are more likely to be applicable to simple datasets in which the dependencies within or between baskets are not so complex. Furthermore, although simple, they are sometimes quite effective for NBR tasks. For exampled, the TIFUKNN \cite{hu2020modeling}, a K-nearest neighbour (KNN) based approach, can even outperform some deep learning based approaches like Dream and Beacon in terms of most of the metrics on the three experimental datasets (cf. TABLE~\ref{tab:statistics_datasets}).


In contrast, deep neural network based approaches are relatively sophisticated, which utilize various neural architectures to capture the intrinsic features and dependencies within and between baskets. As a result, they are usually more time consuming. In some cases, they may not perform as stable as the conventional approaches do. In most cases, deep learning based approaches can better capture the complex intra- and inter-basket dependencies and thus can achieve better recommendation performance. For example, recurrent neural networks (RNN) have been verified as an effective solution in handling sequences of baskets in recent years \cite{r1,r16,r12,r13}.

Last but not least, latent representation based approaches usually do not employ complex deep neural architecture, and thus they are often more efficient and simpler compared with deep learning based approaches. On the other hand, compared with conventional approaches, due to the power of utilized embedding techniques, latent representation based approaches are easier to capture some implicit and hidden patterns in the data and thus are likely to achiever better performance. A typical example in this class is HRM \cite{r15}, which is a distributed representation based NBR approach. It employs a three-layer structure to construct a representation of a user's last basket to predict his/her next basket.


In order to have a better understanding of the representative NBR approaches from different classes and how each category of approaches contribute to promote the evolution of NBR, we systematically compare the selected representative NBR approaches from various classes. We analyze the pros and cons of each approach, which are described in Table \ref{tab:comparison_different_category}.    


\section{Datasets and Compared Approaches}

\begin{table*}[htbp]
\vspace{-20pt}
			\centering
			\caption{A comparison of different classes of NBR approaches.}
			\label{tab:comparison_different_category}
			\resizebox{\linewidth}{!}{
				\begin{tabular}{llp{7cm}p{7cm}}
					\toprule
					\textbf{Approach} & \textbf{Class} & \textbf{Pros} & \textbf{Cons}\\
					\midrule
					\multirow{2}*{TBP \cite{r10}} & \multirow{2}*{Pattern mining} & TBP captures different factors influencing user's next choices. & TBP is usually biased to frequent items while ignoring less-frequent ones.
					\\
					\cline{3-4}
					\multirow{2}*{TIFUKNN \cite{hu2020modeling}}& \multirow{2}*{KNN} & TIFUKNN exploits personalized item frequency information.  & \multirow{4}*{\makecell[l]{They only recommend items that a user has \\ purchased in the past.}} 
					\\
					\cline{3-3}
					\multirow{2}*{UP-CF@r \cite{r14}}& \multirow{2}*{KNN} & UP-CF@r considers the recency of items in the purchased history. & 
					\\
					\cline{3-4}
					\multirow{2}*{FPMC \cite{r2}}& \multirow{2}*{MC} & FPMC models user’s long-term preferences and the transition patterns of items.  & FPMC only captures the first-order dependencies while ignoring the higher-order ones.
					\\   \bottomrule
					\multirow{2}*{NN-Rec \cite{r5}}& \multirow{3}*{Representation} & \multirow{3}*{\makecell[l]{They predict next basket based on the \\ representations of users and baskets.}} & \multirow{3}*{\makecell[l]{
					They fail to capture higher-order dependencies \\ among baskets.}} \\
					\\
					HRM \cite{r15} \\
					\bottomrule
					\multirow{2}*{DREAM \cite{r1}}& \multirow{13}*{RNN}  &DREAM learns a dynamic interests of a user and captures global sequential features of all baskets. & \multirow{13}*{\makecell[l]{They may generate false dependencies and are \\ hard to capture item frequency features.}}
					\\
					\cline{3-3}
					\multirow{2}*{ANAM \cite{r16}}&     &ANAM utilizes the attention mechanism to integrate items and their category information. & 
					\\
					\cline{3-3}
					\multirow{2}*{Beacon \cite{r13}}&   &Beacon utilizes correlation information over items. &
					\\
					\cline{3-3}
					\multirow{2}*{Sets2Sets \cite{r12}}& & Sets2Sets employs an encoder-decoder architecture with repeated purchase pattern. & 
					\\
					\cline{3-3}
					\multirow{2}*{\makecell[l]{IntNet \cite{wang2020intention} \\ Int2Ba \cite{wang2021intention2basket}}} & &\multirow{2}*{\makecell[l]{They consider the human’s intentions contained \\in the purchased  history.}}&\\
					\\
					\cline{3-3}
					\multirow{2}*{CLEA \cite{qin2021world}}&     & CLEA denoises baskets and extracts credibly relevant items to enhance NBR.  &
					\\
					\bottomrule
				\end{tabular}
			}
\end{table*}

\subsection{Datasets}
Through the analysis of datasets in the collected papers, we find two major issues: (1) Due to user privacy limitation, some datasets are not publicly available; (2) For some datasets, though they share the same name, they have different versions actually. For example, we find more than two versions for JingDong dataset, which is provided by the recommender system related competition held by JingDong Retail Group, and is updated each year. Overall, as the purposes and requirements of collected papers are different, different datasets are adopted. 
Figure \ref{fig: summary_four_child}(b) shows the popularity of top-6 datasets, where non-public datasets are not counted. Considering the popularity and characteristics of the datasets above, we select 3 datasets in our experiments, which are commonly used to evaluate the performance of next-basket prediction \cite{wang2020intention}, described as follows:

\begin{itemize}
\item \textbf{TaFeng}\footnote{https://www.kaggle.com/chiranjivdas09/ta-feng-grocery-dataset} released on Kaggle, which contains 4 months of shopping transactions with 32,266 users and 23,812 items. This dataset is a Chinese grocery store dataset with numerous baskets of purchased items. All data in the dataset is utilized in our experiments.

\item \textbf{Instacart}\footnote{https://www.kaggle.com/c/instacart-market-basket-analysis/data} released on Instacart challenge, which contains over 3 million online purchases from more than 200,000 users. Following the work of Qin et al. \cite{qin2021world}, we conduct our experiments by randomly sampling 10\% of the user transaction records from the test user set.

\item \textbf{Dunnhumby}\footnote{https://www.dunnhumby.com/source-files/} released by Dunnhumby, a business data processing and analysis company. It records more than 2 years of purchases of 2,500 households who are frequent shoppers at a retailer. Following the work of Yu et al. \cite{r21}, we adopt the transactions in the first two months to conduct our experiments.
\end{itemize}

The above datasets are publicly available with the information about which user purchases which item at which time, which are suitable to be adopted as benchmark datasets on the task of NBR.

\subsection{Data Pre-processing}



As the original datasets are always huge and sparse, data pre-processing is necessary for the following evaluations. Generally, the infrequent items and inactive users with fewer interactions will be removed during data-processing procedures. By analyzing the collected papers, we find that all papers adopt pre-processing strategies while the strategies they adopt are not uniform. To better fairly evaluate the performance of different approaches across different data, we pre-process the datasets and filter out the items that were purchased less than $n$ times. The values of $n$ are set to 10, 20 and 17 for TaFeng, Instacart, and Dunnhumby datasets, respectively. 
For different datasets, we construct baskets according to its data characteristics. In TaFeng, it contains the timestamp of shopping transactions from 2020-11-01 to 2020-02-28, so we adopt one day as time unit, i.e., the items in the same day count as a basket. In Instacart, we treat all items purchased in the same order as a basket. Similarly, items purchased in the same transaction are treated as a basket in Dunnhumby. For each user, we arrange his/her baskets in chronological order to form a sequence. We only reserve the baskets with the size between 2 to 60 in each dataset. The statistic information of all the datasets after pre-processing is shown in Table \ref{tab:statistics_datasets}. 

Following the work of Hu et al. \cite{hu2020modeling}, we partition the data into 5 folds across sequences, 4 folds is utilized for training, and 1 fold is utilized for testing. We further reserve the data of 10\% sequences in the training set as the validation set to tune hyper-parameters.

\begin{table}[htbp]
 \centering
 \caption{The statistics of experimental datasets.}
 \label{tab:statistics_datasets}
  \begin{tabular}{llll}
  \toprule
  \textbf{Statistics} & \textbf{TaFeng} & \textbf{Instacart} & \textbf{Dunnhumby}\\
  \midrule
  \#Users & 20,212 & 19,982 & 22,530\\
  \#Baskets & 105,140 & 280,941 & 214,861\\
  \#Items & 10,411 & 13,400 & 3,920\\
  \#Basket/user & 5.20 & 14.06 & 9.53\\
  \#Items/basket & 6.32 & 9.61 & 7.45\\
  \bottomrule
 \end{tabular}
 \begin{flushleft}
 \footnotesize
  The rows \#Users, \#Baskets, \#Items, \#Basket/user, \#Items/basket correspond to the number of users, the number of baskets over all users, the number of items, the average number of baskets per user and the average number of  items per basket, respectively.
  \end{flushleft}
\vspace{-22pt}
\end{table}

\subsection{Compared Approaches}

In our collected papers, the compared approaches adopted by each paper are not same. We summarize the top-11 widely-compared approaches, as shown in Figure \ref{fig: summary_four_child}(c). Furthermore, considering their reproducibility and popularity, we select 8 approaches in our experiments. Among them, Sets2Sets is an approach for sequential set prediction with an encoder-decoder framework, which is modified to be applicable for NBR. The details of the eight approaches are described as follows:

\begin{itemize}
\item \textbf{TOP}: a non-personalized approach which recommends the most popular items according to their purchase frequencies \cite{r4}. 

\item \textbf{FPMC}: a Markov Chain (MC) based approach which combines MC with Matrix Factorization (MF) to model the pairwise item-item transition patterns from adjacent baskets to recommend the next basket of items  \cite{r2}.

\item \textbf{DREAM}: an recurrent neural network (RNN) based approach to consider both user's dynamic interest and global sequential features that reflect interactions among baskets  \cite{r1}. 

\item \textbf{Beacon}: an RNN based approach which encodes the basket sequence to model the pairwise correlations among items \cite{r13}.

\item \textbf{Sets2Sets}: an RNN based encoder-decoder approach for set/basket prediction. In addition, an attention mechanism focusing on item frequency is proposed to improve the performance  \cite{r12}.

\item \textbf{UP-CF@r}: a simple approach that relies on the user-wise popularity with collaborative filtering and the recency of shopping \cite{r14}.

\item \textbf{TIFUKNN}: a K-nearest neighbour (KNN) based approach that exploits personalized frequency information of items. A novel repeated purchase pattern is applied in this approach  \cite{hu2020modeling}.

\item \textbf{CLEA}: an RNN based contrastive learning approach to denoise basket generation by identifying relevant items in the history  \cite{qin2021world}.
\end{itemize}

\begin{table*}
\caption{The statistics of compared approaches.}
\label{table: statisticsofallmehtods}
\begin{center}
\setlength{\tabcolsep}{4.0mm
\begin{tabular}{cccccccccc}
\toprule[2pt]
\multicolumn{3}{l}{\multirow{1}{*}{Dataset}} &
 \multicolumn{5}{c}{\multirow{1}{*}{TaFeng}}& \\ \hline
\multicolumn{1}{c}{$K$} &
  \multicolumn{2}{l}{Approaches} &
  \multicolumn{1}{c}{Recall} &
  \multicolumn{1}{c}{Precision} &
  \multicolumn{1}{c}{F1-score} &
  \multicolumn{1}{c}{PHR} &
  \multicolumn{1}{c}{MAP} &
  \multicolumn{1}{c}{MRR} &
  \multicolumn{1}{c}{NDCG} \\ \hline
\multicolumn{1}{c}{\multirow{8}{*}{5}} &
  \multicolumn{2}{l}{TOP} &
  \multicolumn{1}{c}{0.0670} &
  \multicolumn{1}{c}{0.0420} &
  \multicolumn{1}{c}{0.0412} &
  \multicolumn{1}{c}{0.1870} &
  \multicolumn{1}{c}{0.1426} &
  \multicolumn{1}{c}{0.1438} &
  \multicolumn{1}{c}{0.0733} \\ \cline{2-10} 
\multicolumn{1}{c}{} &
  \multicolumn{2}{l}{FPMC} &
  \multicolumn{1}{c}{0.0632} &
  \multicolumn{1}{c}{0.0433} &
  \multicolumn{1}{c}{0.0514} &
  \multicolumn{1}{c}{0.1912} &
  \multicolumn{1}{c}{0.1066} &
  \multicolumn{1}{c}{0.1070} &
  \multicolumn{1}{c}{0.0557} \\ \cline{2-10} 
\multicolumn{1}{c}{} &
  \multicolumn{2}{l}{DREAM} &
  \multicolumn{1}{c}{0.0921} &
  \multicolumn{1}{c}{0.0533} &
  \multicolumn{1}{c}{0.0675} &
  \multicolumn{1}{c}{0.2329} &
  \multicolumn{1}{c}{0.1580} &
  \multicolumn{1}{c}{0.1608} &
  \multicolumn{1}{c}{0.0891} \\ \cline{2-10} 
\multicolumn{1}{c}{} &
  \multicolumn{2}{l}{Beacon} &
  \multicolumn{1}{c}{0.0820} &
  \multicolumn{1}{c}{0.0489} &
  \multicolumn{1}{c}{0.0493} &
  \multicolumn{1}{c}{0.2117} &
  \multicolumn{1}{c}{0.1473} &
  \multicolumn{1}{c}{0.1510} &
  \multicolumn{1}{c}{0.0817} \\ \cline{2-10}
\multicolumn{1}{c}{} &
  \multicolumn{2}{l}{Sets2Sets} &
  \multicolumn{1}{c}{0.0951} &
  \multicolumn{1}{c}{0.0690} &
  \multicolumn{1}{c}{0.0647} &
  \multicolumn{1}{c}{0.2860} &
  \multicolumn{1}{c}{0.1718} &
  \multicolumn{1}{c}{0.1763} &
  \multicolumn{1}{c}{0.0925} \\ \cline{2-10}
\multicolumn{1}{c}{} &
  \multicolumn{2}{l}{UP-CF@r} &
  \multicolumn{1}{c}{0.0741} &
  \multicolumn{1}{c}{0.0644} &
  \multicolumn{1}{c}{0.0565} &
  \multicolumn{1}{c}{0.2601} &
  \multicolumn{1}{c}{0.1539} &
  \multicolumn{1}{c}{0.1574} &
  \multicolumn{1}{c}{0.0726} \\ \cline{2-10} 
\multicolumn{1}{c}{} &
   \multicolumn{2}{l}{TIFUKNN} &
  \multicolumn{1}{c}{0.0833} &
  \multicolumn{1}{c}{0.0713} &
  \multicolumn{1}{c}{0.0624} &
  \multicolumn{1}{c}{0.2799} &
  \multicolumn{1}{c}{0.1626} &
  \multicolumn{1}{c}{0.1663} &
  \multicolumn{1}{c}{0.0790} \\ \cline{2-10}
\multicolumn{1}{c}{} &
  \multicolumn{2}{l}{CLEA} &
  \multicolumn{1}{c}{\textbf{0.1280}} &
  \multicolumn{1}{c}{\textbf{0.0900}} &
  \multicolumn{1}{c}{\textbf{0.0842}} &
  \multicolumn{1}{c}{\textbf{0.3547}} &
  \multicolumn{1}{c}{\textbf{0.2321}} &
  \multicolumn{1}{c}{\textbf{0.2386}} &
  \multicolumn{1}{c}{\textbf{0.1284}} \\ 
  \hline
\multicolumn{1}{c}{\multirow{7}{*}{10}} &
  \multicolumn{2}{l}{TOP} &
  \multicolumn{1}{c}{0.0768} &
  \multicolumn{1}{c}{0.0269} &
  \multicolumn{1}{c}{0.0326} &
  \multicolumn{1}{c}{0.2282} &
  \multicolumn{1}{c}{0.1455} &
  \multicolumn{1}{c}{0.1495} &
  \multicolumn{1}{c}{0.0786} \\ \cline{2-10} 
\multicolumn{1}{c}{} &
  \multicolumn{2}{l}{FPMC} &
  \multicolumn{1}{c}{0.0755} &
  \multicolumn{1}{c}{0.0274} &
  \multicolumn{1}{c}{0.0402} &
  \multicolumn{1}{c}{0.2314} &
  \multicolumn{1}{c}{0.1096} &
  \multicolumn{1}{c}{0.1122} &
  \multicolumn{1}{c}{0.0616} \\ \cline{2-10} 
\multicolumn{1}{c}{} &
  \multicolumn{2}{l}{DREAM} &
  \multicolumn{1}{c}{0.1170} &
  \multicolumn{1}{c}{0.0362} &
  \multicolumn{1}{c}{0.0553} &
  \multicolumn{1}{c}{0.3060} &
  \multicolumn{1}{c}{0.1648} &
  \multicolumn{1}{c}{0.1705} &
  \multicolumn{1}{c}{0.1004} \\ \cline{2-10} 
\multicolumn{1}{c}{} &
  \multicolumn{2}{l}{Beacon} &
  \multicolumn{1}{c}{0.1059} &
  \multicolumn{1}{c}{0.0348} &
  \multicolumn{1}{c}{0.0435} &
  \multicolumn{1}{c}{0.2863} &
  \multicolumn{1}{c}{0.1543} &
  \multicolumn{1}{c}{0.1608} &
  \multicolumn{1}{c}{0.0930} \\ \cline{2-10} 
\multicolumn{1}{c}{} &
  \multicolumn{2}{l}{Sets2Sets} &
  \multicolumn{1}{c}{0.1315} &
  \multicolumn{1}{c}{0.0512} &
  \multicolumn{1}{c}{0.0614} &
  \multicolumn{1}{c}{0.3797} &
  \multicolumn{1}{c}{0.1730} &
  \multicolumn{1}{c}{0.1816} &
  \multicolumn{1}{c}{0.1082} \\ \cline{2-10} 
\multicolumn{1}{c}{} &
  \multicolumn{2}{l}{UP-CF@r} &
  \multicolumn{1}{c}{0.1058} &
  \multicolumn{1}{c}{0.0496} &
  \multicolumn{1}{c}{0.0560} &
  \multicolumn{1}{c}{0.3483} &
  \multicolumn{1}{c}{0.1600} &
  \multicolumn{1}{c}{0.1691} &
  \multicolumn{1}{c}{0.0889} \\ \cline{2-10} 
\multicolumn{1}{c}{} &
  \multicolumn{2}{l}{TIFUKNN} &
  \multicolumn{1}{c}{0.1236} &
  \multicolumn{1}{c}{0.0537} &
  \multicolumn{1}{c}{0.0619} &
  \multicolumn{1}{c}{0.3765} &
  \multicolumn{1}{c}{0.1680} &
  \multicolumn{1}{c}{0.1791} &
  \multicolumn{1}{c}{0.0982} \\
  \cline{2-10}
\multicolumn{1}{c}{} &
  \multicolumn{2}{l}{CLEA} &
  \multicolumn{1}{c}{\textbf{0.1609}} &
  \multicolumn{1}{c}{\textbf{0.0648}} &
  \multicolumn{1}{c}{\textbf{0.0760}} &
  \multicolumn{1}{c}{\textbf{0.4446}} &
  \multicolumn{1}{c}{\textbf{0.2353}} &
  \multicolumn{1}{c}{\textbf{0.2514}} &
  \multicolumn{1}{c}{\textbf{0.1472}}\\ 
 \toprule[2pt]
\multicolumn{3}{l}{\multirow{1}{*}{Dataset}} &
\multicolumn{5}{c}{\multirow{1}{*}{Instacart}}& \\ \hline
\multicolumn{1}{c}{$K$} &
  \multicolumn{2}{l}{Approaches} &
  \multicolumn{1}{c}{Recall} &
  \multicolumn{1}{c}{Precision} &
  \multicolumn{1}{c}{F1-score} &
  \multicolumn{1}{c}{PHR} &
  \multicolumn{1}{c}{MAP} &
  \multicolumn{1}{c}{MRR} &
  \multicolumn{1}{c}{NDCG} \\ \hline
\multicolumn{1}{c}{\multirow{8}{*}{5}} &
  \multicolumn{2}{l}{TOP} &
  \multicolumn{1}{c}{0.0487} &
  \multicolumn{1}{c}{0.0956} &
  \multicolumn{1}{c}{0.0581} &
  \multicolumn{1}{c}{0.3668} &
  \multicolumn{1}{c}{0.2270} &
  \multicolumn{1}{c}{0.2321} &
  \multicolumn{1}{c}{0.0666} \\ \cline{2-10} 
\multicolumn{1}{c}{} &
  \multicolumn{2}{l}{FPMC} &
  \multicolumn{1}{c}{0.0481} &
  \multicolumn{1}{c}{0.0948} &
  \multicolumn{1}{c}{0.0620} &
  \multicolumn{1}{c}{0.3600} &
  \multicolumn{1}{c}{0.2211} &
  \multicolumn{1}{c}{0.2264} &
  \multicolumn{1}{c}{0.0651} \\ \cline{2-10} 
\multicolumn{1}{c}{} &
  \multicolumn{2}{l}{DREAM} &
  \multicolumn{1}{c}{0.0763} &
  \multicolumn{1}{c}{0.0486} &
  \multicolumn{1}{c}{0.0594} &
  \multicolumn{1}{c}{0.2176} &
  \multicolumn{1}{c}{0.1543} &
  \multicolumn{1}{c}{0.1481} &
  \multicolumn{1}{c}{0.0754} \\ \cline{2-10} 
\multicolumn{1}{c}{} &
  \multicolumn{2}{l}{Beacon} &
  \multicolumn{1}{c}{0.0539} &
  \multicolumn{1}{c}{0.1049} &
  \multicolumn{1}{c}{0.0638} &
  \multicolumn{1}{c}{0.3930} &
  \multicolumn{1}{c}{0.2181} &
  \multicolumn{1}{c}{0.2250} &
  \multicolumn{1}{c}{0.0695} \\ \cline{2-10}
\multicolumn{1}{c}{} &
  \multicolumn{2}{l}{Sets2Sets} &
  \multicolumn{1}{c}{0.1266} &
  \multicolumn{1}{c}{0.1652} &
  \multicolumn{1}{c}{0.1209} &
  \multicolumn{1}{c}{0.5503} &
  \multicolumn{1}{c}{0.3141} &
  \multicolumn{1}{c}{0.3260} &
  \multicolumn{1}{c}{0.1375} \\ \cline{2-10}
\multicolumn{1}{c}{} &
  \multicolumn{2}{l}{UP-CF@r} &
  \multicolumn{1}{c}{0.2512} &
  \multicolumn{1}{c}{0.3580} &
  \multicolumn{1}{c}{0.2512} &
  \multicolumn{1}{c}{0.7946} &
  \multicolumn{1}{c}{0.5818} &
  \multicolumn{1}{c}{0.6138} &
  \multicolumn{1}{c}{0.2970} \\ \cline{2-10} 
\multicolumn{1}{c}{} &
   \multicolumn{2}{l}{TIFUKNN} &
  \multicolumn{1}{c}{\textbf{0.2616}} &
  \multicolumn{1}{c}{\textbf{0.3733}} &
  \multicolumn{1}{c}{\textbf{0.2619}} &
  \multicolumn{1}{c}{\textbf{0.8052}} &
  \multicolumn{1}{c}{\textbf{0.5992}} &
  \multicolumn{1}{c}{\textbf{0.6312}} &
  \multicolumn{1}{c}{\textbf{0.3094}} \\ \cline{2-10}
\multicolumn{1}{c}{} &
  \multicolumn{2}{l}{CLEA} &
  \multicolumn{1}{c}{0.1850} &
  \multicolumn{1}{c}{0.3025} &
  \multicolumn{1}{c}{0.1973} &
  \multicolumn{1}{c}{0.7136} &
  \multicolumn{1}{c}{0.5623} &
  \multicolumn{1}{c}{0.5865} &
  \multicolumn{1}{c}{0.2450} \\ 
  \hline
\multicolumn{1}{c}{\multirow{7}{*}{10}} &
  \multicolumn{2}{l}{TOP} &
  \multicolumn{1}{c}{0.0731} &
  \multicolumn{1}{c}{0.0735} &
  \multicolumn{1}{c}{0.0655} &
  \multicolumn{1}{c}{0.4614} &
  \multicolumn{1}{c}{0.2223} &
  \multicolumn{1}{c}{0.2448} &
  \multicolumn{1}{c}{0.0832} \\ \cline{2-10} 
\multicolumn{1}{c}{} &
  \multicolumn{2}{l}{FPMC} &
  \multicolumn{1}{c}{0.0712} &
  \multicolumn{1}{c}{0.0716} &
  \multicolumn{1}{c}{0.0697} &
  \multicolumn{1}{c}{0.4508} &
  \multicolumn{1}{c}{0.2182} &
  \multicolumn{1}{c}{0.2388} &
  \multicolumn{1}{c}{0.0809} \\ \cline{2-10} 
\multicolumn{1}{c}{} &
  \multicolumn{2}{l}{DREAM} &
  \multicolumn{1}{c}{0.1012} &
  \multicolumn{1}{c}{0.0359} &
  \multicolumn{1}{c}{0.0530} &
  \multicolumn{1}{c}{0.3037} &
  \multicolumn{1}{c}{0.1543} &
  \multicolumn{1}{c}{0.1596} &
  \multicolumn{1}{c}{0.0874} \\ \cline{2-10} 
\multicolumn{1}{c}{} &
  \multicolumn{2}{l}{Beacon} &
  \multicolumn{1}{c}{0.0767} &
  \multicolumn{1}{c}{0.0769} &
  \multicolumn{1}{c}{0.0686} &
  \multicolumn{1}{c}{0.4755} &
  \multicolumn{1}{c}{0.2174} &
  \multicolumn{1}{c}{0.2362} &
  \multicolumn{1}{c}{0.0851} \\ \cline{2-10}
\multicolumn{1}{c}{} &
  \multicolumn{2}{l}{Sets2Sets} &
  \multicolumn{1}{c}{0.2058} &
  \multicolumn{1}{c}{0.1463} &
  \multicolumn{1}{c}{0.1710} &
  \multicolumn{1}{c}{0.7157} &
  \multicolumn{1}{c}{0.3155} &
  \multicolumn{1}{c}{0.3550} &
  \multicolumn{1}{c}{0.1862} \\ \cline{2-10}
\multicolumn{1}{c}{} &
  \multicolumn{2}{l}{UP-CF@r} &
  \multicolumn{1}{c}{0.3480} &
  \multicolumn{1}{c}{0.2731} &
  \multicolumn{1}{c}{0.2650} &
  \multicolumn{1}{c}{0.8590} &
  \multicolumn{1}{c}{0.5448} &
  \multicolumn{1}{c}{0.6226} &
  \multicolumn{1}{c}{0.3613} \\ \cline{2-10} 
\multicolumn{1}{c}{} &
   \multicolumn{2}{l}{TIFUKNN} &
  \multicolumn{1}{c}{\textbf{0.3698}} &
  \multicolumn{1}{c}{\textbf{0.2896}} &
  \multicolumn{1}{c}{\textbf{0.2812}} &
  \multicolumn{1}{c}{\textbf{0.8756}} &
  \multicolumn{1}{c}{\textbf{0.5618}} &
  \multicolumn{1}{c}{\textbf{0.6409}} &
  \multicolumn{1}{c}{\textbf{0.3805}} \\ \cline{2-10}
\multicolumn{1}{c}{} &
  \multicolumn{2}{l}{CLEA} &
  \multicolumn{1}{c}{0.2135} &
  \multicolumn{1}{c}{0.1857} &
  \multicolumn{1}{c}{0.1738} &
  \multicolumn{1}{c}{0.7439} &
  \multicolumn{1}{c}{0.5020} &
  \multicolumn{1}{c}{0.5659} &
  \multicolumn{1}{c}{0.2537} \\ 
 \toprule[2pt]
\multicolumn{3}{l}{\multirow{1}{*}{Dataset}} &
\multicolumn{5}{c}{\multirow{1}{*}{Dunnhumby}}& \\ \hline
\multicolumn{1}{c}{$K$} &
  \multicolumn{2}{l}{Approaches} &
  \multicolumn{1}{c}{Recall} &
  \multicolumn{1}{c}{Precision} &
  \multicolumn{1}{c}{F1-score} &
  \multicolumn{1}{c}{PHR} &
  \multicolumn{1}{c}{MAP} &
  \multicolumn{1}{c}{MRR} &
  \multicolumn{1}{c}{NDCG} \\ \hline
\multicolumn{1}{c}{\multirow{8}{*}{5}} &
  \multicolumn{2}{l}{TOP} &
  \multicolumn{1}{c}{0.0966} &
  \multicolumn{1}{c}{0.0763} &
  \multicolumn{1}{c}{0.0606} &
  \multicolumn{1}{c}{0.3294} &
  \multicolumn{1}{c}{0.2190} &
  \multicolumn{1}{c}{0.2282} &
  \multicolumn{1}{c}{0.0861} \\ \cline{2-10} 
\multicolumn{1}{c}{} &
  \multicolumn{2}{l}{FPMC} &
  \multicolumn{1}{c}{0.0746} &
  \multicolumn{1}{c}{0.1033} &
  \multicolumn{1}{c}{0.0866} &
  \multicolumn{1}{c}{0.3928} &
  \multicolumn{1}{c}{0.2566} &
  \multicolumn{1}{c}{0.2674} &
  \multicolumn{1}{c}{0.0903} \\ \cline{2-10} 
\multicolumn{1}{c}{} &
  \multicolumn{2}{l}{DREAM} &
  \multicolumn{1}{c}{0.0756} &
  \multicolumn{1}{c}{0.1049} &
  \multicolumn{1}{c}{0.0879} &
  \multicolumn{1}{c}{0.3984} &
  \multicolumn{1}{c}{0.2616} &
  \multicolumn{1}{c}{0.2726} &
  \multicolumn{1}{c}{0.0918} \\ \cline{2-10} 
\multicolumn{1}{c}{} &
  \multicolumn{2}{l}{Beacon} &
  \multicolumn{1}{c}{0.0795} &
  \multicolumn{1}{c}{0.1038} &
  \multicolumn{1}{c}{0.0737} &
  \multicolumn{1}{c}{0.3948} &
  \multicolumn{1}{c}{0.2517} &
  \multicolumn{1}{c}{0.2682} &
  \multicolumn{1}{c}{0.0932} \\ \cline{2-10}
\multicolumn{1}{c}{} &
  \multicolumn{2}{l}{Sets2Sets} &
  \multicolumn{1}{c}{0.1040} &
  \multicolumn{1}{c}{0.1249} &
  \multicolumn{1}{c}{0.0904} &
  \multicolumn{1}{c}{0.4353} &
  \multicolumn{1}{c}{0.2515} &
  \multicolumn{1}{c}{0.2867} &
  \multicolumn{1}{c}{0.1112} \\ \cline{2-10}
\multicolumn{1}{c}{} &
  \multicolumn{2}{l}{UP-CF@r} &
  \multicolumn{1}{c}{\textbf{0.1794}} &
  \multicolumn{1}{c}{\textbf{0.2417}} &
  \multicolumn{1}{c}{\textbf{0.1693}} &
  \multicolumn{1}{c}{\textbf{0.6005}} &
  \multicolumn{1}{c}{\textbf{0.4367}} &
  \multicolumn{1}{c}{\textbf{0.4602}} &
  \multicolumn{1}{c}{\textbf{0.2105}} \\ \cline{2-10} 
\multicolumn{1}{c}{} &
   \multicolumn{2}{l}{TIFUKNN} &
  \multicolumn{1}{c}{0.1725} &
  \multicolumn{1}{c}{0.2329} &
  \multicolumn{1}{c}{0.1630} &
  \multicolumn{1}{c}{0.5947} &
  \multicolumn{1}{c}{0.4260} &
  \multicolumn{1}{c}{0.4465} &
  \multicolumn{1}{c}{0.2027} \\ \cline{2-10}
\multicolumn{1}{c}{} &
  \multicolumn{2}{l}{CLEA} &
  \multicolumn{1}{c}{0.1193} &
  \multicolumn{1}{c}{0.1720} &
  \multicolumn{1}{c}{0.1165} &
  \multicolumn{1}{c}{0.5042} &
  \multicolumn{1}{c}{0.3512} &
  \multicolumn{1}{c}{0.3663} &
  \multicolumn{1}{c}{0.1422} \\ 
  \hline
\multicolumn{1}{c}{\multirow{7}{*}{10}} &
  \multicolumn{2}{l}{TOP} &
  \multicolumn{1}{c}{0.1169} &
  \multicolumn{1}{c}{0.0566} &
  \multicolumn{1}{c}{0.0555} &
  \multicolumn{1}{c}{0.4161} &
  \multicolumn{1}{c}{0.2194} &
  \multicolumn{1}{c}{0.2385} &
  \multicolumn{1}{c}{0.0983} \\ \cline{2-10} 
\multicolumn{1}{c}{} &
  \multicolumn{2}{l}{FPMC} &
  \multicolumn{1}{c}{0.0961} &
  \multicolumn{1}{c}{0.0705} &
  \multicolumn{1}{c}{0.0813} &
  \multicolumn{1}{c}{0.4560} &
  \multicolumn{1}{c}{0.2483} &
  \multicolumn{1}{c}{0.2758} &
  \multicolumn{1}{c}{0.1037} \\ \cline{2-10} 
\multicolumn{1}{c}{} &
  \multicolumn{2}{l}{DREAM} &
  \multicolumn{1}{c}{0.1006} &
  \multicolumn{1}{c}{0.0732} &
  \multicolumn{1}{c}{0.0847} &
  \multicolumn{1}{c}{0.4711} &
  \multicolumn{1}{c}{0.2535} &
  \multicolumn{1}{c}{0.2822} &
  \multicolumn{1}{c}{0.1069} \\ \cline{2-10} 
\multicolumn{1}{c}{} &
  \multicolumn{2}{l}{Beacon} &
  \multicolumn{1}{c}{0.1030} &
  \multicolumn{1}{c}{0.0726} &
  \multicolumn{1}{c}{0.0709} &
  \multicolumn{1}{c}{0.4667} &
  \multicolumn{1}{c}{0.2493} &
  \multicolumn{1}{c}{0.2776} &
  \multicolumn{1}{c}{0.1079} \\ \cline{2-10}
\multicolumn{1}{c}{} &
  \multicolumn{2}{l}{Sets2Sets} &
  \multicolumn{1}{c}{0.1695} &
  \multicolumn{1}{c}{0.1102} &
  \multicolumn{1}{c}{0.1091} &
  \multicolumn{1}{c}{0.5751} &
  \multicolumn{1}{c}{0.2563} &
  \multicolumn{1}{c}{0.2867} &
  \multicolumn{1}{c}{0.1488} \\ \cline{2-10}
\multicolumn{1}{c}{} &
  \multicolumn{2}{l}{UP-CF@r} &
  \multicolumn{1}{c}{\textbf{0.2480}} &
  \multicolumn{1}{c}{\textbf{0.1795}} &
  \multicolumn{1}{c}{\textbf{0.1733}} &
  \multicolumn{1}{c}{\textbf{0.6764}} &
  \multicolumn{1}{c}{\textbf{0.4149}} &
  \multicolumn{1}{c}{\textbf{0.4703}} &
  \multicolumn{1}{c}{\textbf{0.2525}} \\ \cline{2-10} 
\multicolumn{1}{c}{} &
   \multicolumn{2}{l}{TIFUKNN} &
  \multicolumn{1}{c}{0.2419} &
  \multicolumn{1}{c}{0.1734} &
  \multicolumn{1}{c}{0.1677} &
  \multicolumn{1}{c}{0.6713} &
  \multicolumn{1}{c}{0.4054} &
  \multicolumn{1}{c}{0.4568} &
  \multicolumn{1}{c}{0.2444} \\ \cline{2-10}
\multicolumn{1}{c}{} &
  \multicolumn{2}{l}{CLEA} &
  \multicolumn{1}{c}{0.1476} &
  \multicolumn{1}{c}{0.1102} &
  \multicolumn{1}{c}{0.1048} &
  \multicolumn{1}{c}{0.5555} &
  \multicolumn{1}{c}{0.3555} &
  \multicolumn{1}{c}{0.3861} &
  \multicolumn{1}{c}{0.1673} \\ 
  \toprule[2pt]
\end{tabular}
 }
\end{center}
\end{table*}

\begin{figure*} 
    \vspace{-15pt}
    \centering
        \subfigure[TaFeng]{
        \includegraphics[width=0.31\linewidth]{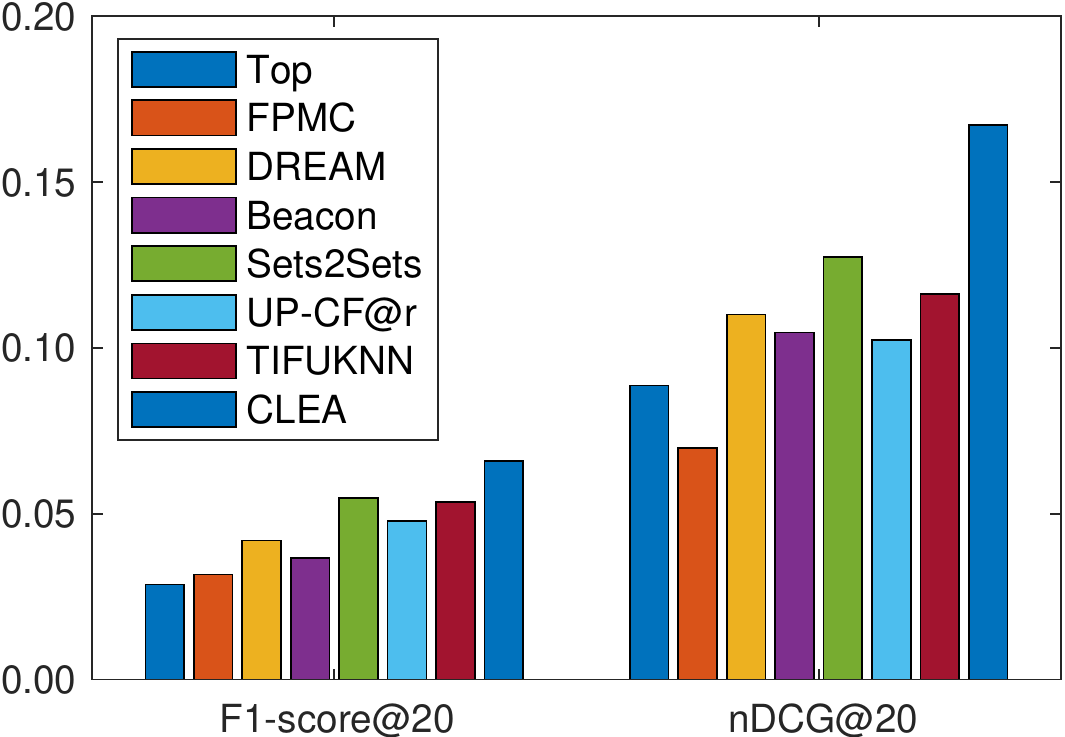}
        }
        \subfigure[Instacart]{
        \includegraphics[width=0.31\linewidth]{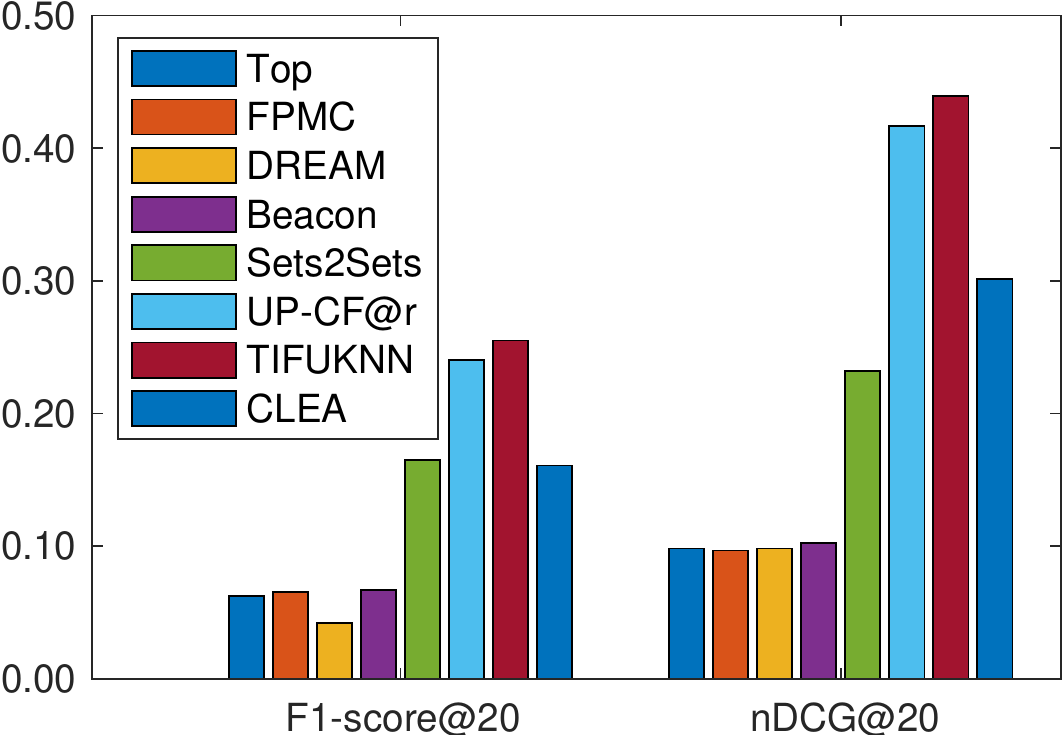}
        }
        \subfigure[Dunnhumby]{
        \includegraphics[width=0.31\linewidth]{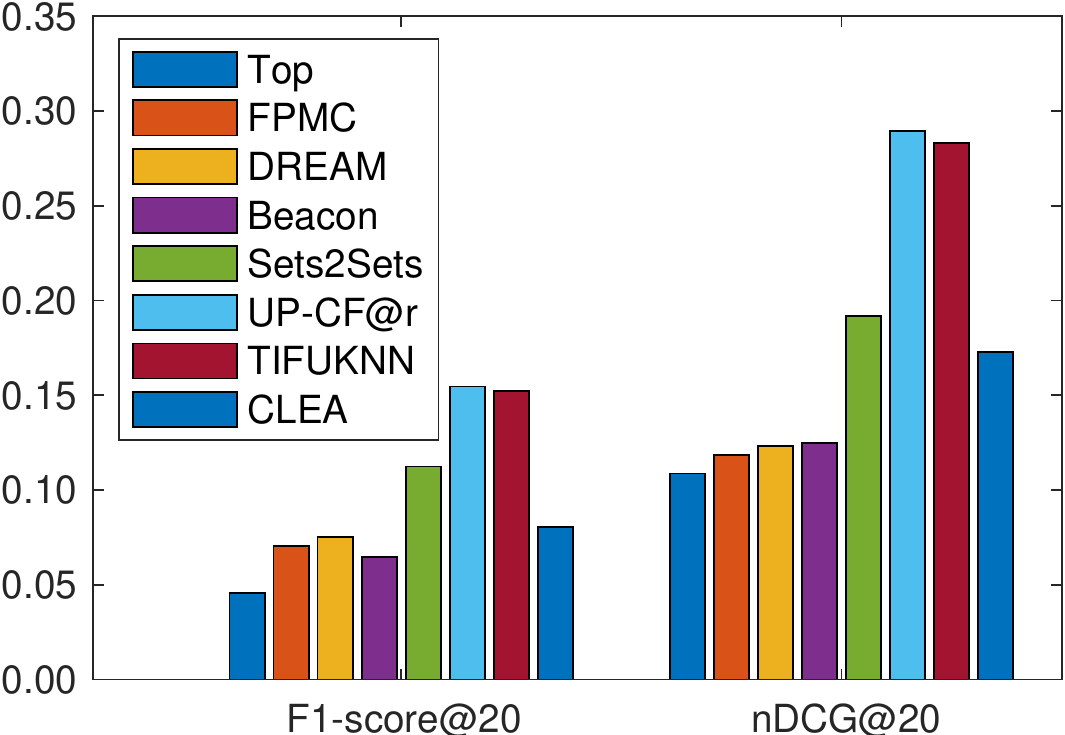}
        }
    \caption{
    The performance of different approaches in terms of F1-Score and NDCG.
     }

	  \label{fig:performaneF1Ndcg} 
\end{figure*}

\section{Benchmarking Evaluation}

\subsection{Evaluation Metrics}
In our collected papers, the evaluation metrics adopted by each paper change greatly. Figure \ref{fig: summary_four_child}(d) shows the popularity of evaluation metrics. In order to comprehensively and fairly evaluate different approaches, we adopt the following 7 metrics: Recall, Precision, F1-Score, Person-wise Hit Ratio (PHR), Normalized Discounted Cumulative Gain (NDCG), Mean Average Precision (MAP) and Mean Reciprocal Rank (MRR), where the latter two metrics are not reflected in the collected papers, but they are common in the top-$K$ recommendation. The details of the seven metrics are introduced as follows:
\begin{itemize}
\item \textbf{Recall}: it is a widely-used metric in NBR \cite{r20}, which measures the proportion of ground-truth items in a predicted basket that are correctly recommended. Recall is calculated by: 

\[{\rm{Recall@K}} = \frac{{\left| {S' \cap S} \right|}}{{\left| S \right|}},\]

where $S$ and $S'$ are the ground-truth items and the top-$K$ items of predicted basket respectively,  $\lvert S \rvert$ denotes the size of ground-truth items. 

\item \textbf{Precision}: it is corresponding with Recall, measures the proportion of ground-truth items in the top-$K$ items of predicted baskets \cite{r20},  which is calculated by: 
\[{\rm{Precision@K}} = \frac{{\left| {S' \cap S} \right|}}{K}.\]

\item \textbf{F1-Score}: it is the harmonic mean of precision and recall \cite{r15}, which is calculated by:
 \[\rm{F1}-\rm{Score@K} =  \frac{{2 \times Precision@K \times Recall@K}}{{\ Precision@K + Recall@K}}.\]

\item \textbf{PHR}: it is person-wise hit ratio which represents the ratio of users whose ground-truth items appear in the recommendation list\cite{r21}. PHR is calculated by

\[{\rm{PHR@K}} = \frac{1}{N}\sum\limits_{i = 1}^N {hr(i)}, \]where

\[hr(i) = \left\{ \begin{array}{l}
1,\quad \left| {S' \cap S} \right| > 0\\
0,\quad \left| {S' \cap S} \right| = 0,
\end{array} \right.\] and $N$ denotes the number of testing users.


\item \textbf{NDCG}: it is a ranking based measure, which focuses on the order of items in a predicted basket \cite{r21}. NDCG is more sensitive to higher ranked items. That is to say, higher NDCG indicates the ground-truth items are recommended at higher ranks. NDCG is defined as:

\[{\rm{NDCG@K   =   }}\frac{1}{{\sum\limits_{j = 1}^{\left| S \right|} {\frac{1}{{{{\log }_2}(j + 1)}}} }}{\rm{ }}\sum\limits_{i = 1}^K {\frac{{\partial (S',S)}}{{{{\log }_2}(k + 1)}}} ,\]

where $\partial (S',S)$ returns 1 when the item in the predicted basket appears in the ground-truth, otherwise 0.

\item  \textbf{MAP}: it is a relatively common evaluation metric \cite{r4}. Average Precision (AP) is calculated in the following way:

\[{\rm{AP}}(i) = \frac{1}{m}\sum\limits_{i = 1}^m {m \times \frac{1}{{{p_i}}}},\]

where $m$ is the number of ground-truth items if they appear in the recommendation list ($m<= K$), and ${p_i}$ is the position of item $i$ in the recommendation list. And, MAP is the mean of AP, which is calculated by: 

\[{\rm{MAP@K}} = \frac{1}{N}\sum\limits_{i = 1}^N {AP(i)}, \]

where $N$ is same to that in PHR.

\item \textbf{MRR}: it is derived from the information retrieval. Reciprocal Rank(RR) refers to the inverse of the ranking of correctly recommended item in the top-$K$ recommendation list \cite{r4}. MRR is the mean of Reciprocal Rank, which is calculated by:
\[{\rm{MRR@K}} = \frac{1}{N}\sum\limits_{i = 1}^N {\frac{1}{{{q_i}}}},\]
where we define ${q_i}$ as the ranking of the first item that appears in the ground-truth in the top-$K$ recommendation list.  
\end{itemize}

Following the settings of traditional NBR work \cite{hu2020modeling,qin2021world}, for each approach, we will recommend a predicted basket (i.e., recommendation list) with a fixed size $K$ for evaluation. All the metrics are calculated across all predicted baskets, and all metrics have the same characteristic: the larger value, the better performance.

\begin{table*}[htbp]
\vspace{-10pt}
\caption{A list of representative open-source NBRs algorithms}
\label{table: statisticsofallalgorithms}
\begin{center}
\setlength{\tabcolsep}{2.3mm
\begin{tabular}{|l|l|l|l|ll}
\cline{1-4}
\textbf{Algorithm} & \multicolumn{1}{c|}{\textbf{Utilized model}}                                                         & \textbf{Venue}                   & \multicolumn{1}{c|}{\textbf{Link}}                                                                 &  &  \\ \cline{1-4}
DREAM\cite{r1}              & RNN                                                                                                  & SIGIR 2016                       & \href{https://github.com/yihong-chen/DREAM}{https://github.com/yihong-chen/DREAM}                                                               &  &  \\ \cline{1-4}
Sets2Sets\cite{r12}          & RNN                                                                                                  & KDD 2019                         & \href{https://github.com/HaojiHu/Sets2Sets}{https://github.com/HaojiHu/Sets2Sets}                                                                &  &  \\ \cline{1-4}
Beacon\cite{r13}             & RNN                                                                                                  & IJCAI 2019                       & \href{https://github.com/PreferredAI/beacon}{https://github.com/PreferredAI/beacon}                                                              &  &  \\ \cline{1-4}
CLEA\cite{qin2021world}               & RNN                                                                                                  & SIGIR 2021                       & \href{https://github.com/QYQ-bot/CLEA}{https://github.com/QYQ-bot/CLEA}                                                                    &  &  \\ \cline{1-4}
MBN\cite{shen2022mbn}                & RNN                                                                                                  & TKDD 2022 & \href{https://github.com/gybuay/MBN}{https://github.com/gybuay/MBN}                                                                      &  &  \\ \cline{1-4}
HRM\cite{r15}                & Distributed representation                             & SIGIR 2015                       & \href{https://github.com/chenghu17/Sequential_Recommendation}{https://github.com/chenghu17/Sequential\_Recommendation} &  &  \\ \cline{1-4}
ReCANet\cite{ariannezhad2022recanet}            & Distributed representation & SIGIR 2022                       & \href{https://github.com/mzhariann/recanet}{https://github.com/mzhariann/recanet}                                                               &  &  \\ \cline{1-4}
UP-CF@r\cite{r14}            & KNN                                                                                                  & UMAP 2020                        & \href{https://github.com/MayloIFERR/RACF}{https://github.com/MayloIFERR/RACF}                                                                 &  &  \\ \cline{1-4}
TIFUKNN\cite{hu2020modeling}            & KNN                                                                                                  & SIGIR 2020                       & \href{https://github.com/HaojiHu/TIFUKNN}{https://github.com/HaojiHu/TIFUKNN}                                                                 &  &  \\ \cline{1-4}
DNNTSP\cite{r21}             & GNN                                                                                                  & KDD 2020                         & \href{https://github.com/yule-BUAA/DNNTSP}{https://github.com/yule-BUAA/DNNTSP}                                                                &  &  \\ \cline{1-4}
MITGNN\cite{Liu2020MITGNN}             & GNN                                                                                                  & BigData 2020                     & \href{https://github.com/JimLiu96/MITGNN}{https://github.com/JimLiu96/MITGNN}                                                                 &  &  \\ \cline{1-4}
FPMC\cite{r2}               & MC                                                                                                   & WWW 2010                         & \href{https://github.com/khesui/FPMC}{https://github.com/khesui/FPMC}                                                                     &  &  \\ \cline{1-4}
TBP\cite{r10}                & Pattern   mining                                                                                     & ICDM 2017                        & \href{https://github.com/GiulioRossetti/tbp-next-basket}{https://github.com/GiulioRossetti/tbp-next-basket}                                                  &  &  \\ \cline{1-4}
\end{tabular}
}
\end{center}
\end{table*}

\begin{table*}[htbp]
\caption{A list of commonly used and publicly accessible real-world datasets for NBRs}
\label{table: statisticsofalldatasets}
\begin{center}
 \begin{threeparttable}
\setlength{\tabcolsep}{1.0mm
\begin{tabular}{|l|l|l|l|l|l|l|}
\hline
\textbf{Dataset}       & \textbf{\#Users}   & \textbf{\#Items}  & \textbf{\#Baskets} & \textbf{Avg. basket size} & \textbf{\#Baskets per user} & \textbf{Reference} \\ \hline
TaFeng        & 12,805  & 10,829 & 89,543        & 6.39         & 6.99  & \cite{r15}\cite{r1}\cite{r10}\cite{r16}\cite{r12}\cite{r13}\cite{qin2021world}\cite{hu2020modeling}\cite{r21}          \\ \hline
Instacart     & 7,282   & 12,515 & 115,717       & 9.63         & 15.89 & \cite{r14}\cite{wang2020modelling}\cite{Liu2020MITGNN}\cite{qin2021world}\cite{ariannezhad2022recanet}          \\ \hline
Dunnhumby     & 2,488   & 26,779 & 269,951      & 9.02         & 108.50 &\cite{r12}\cite{ariannezhad2022recanet}\cite{qin2021world}\cite{hu2020modeling}\cite{r14}           \\ \hline
Tmall\tnote{5}       & 102,681 & 36,113 & 739,178        & 1.72         & 7.19 &\cite{r15}\cite{r1}\cite{wang2021intention2basket}\cite{wang2020intention}\cite{shen2022mbn}          \\ \hline
JingDong\tnote{6}      & 60,534  & 41,186 & 243,769        & 1.16         & 4.02 & \cite{r16}\cite{shen2022mbn}          \\ \hline
Tianchi\tnote{7}     & 6,924   & 27,637 & 34,749        & 2.12         & 5.01  &   \cite{shen2022mbn}         \\ \hline
Valuedshopper\tnote{8} & 9,997   & 6,421  & 280,762       & 9.17         & 28.08 & \cite{qin2021world}\cite{ariannezhad2022recanet}\cite{hu2020modeling}          \\ \hline
Taobao\tnote{9}       & 47,392  & 90,440 & 169,840         & 1.48         & 3.58 &  \cite{r21}         \\ \hline
\end{tabular}
}
      \begin{tablenotes}
        \footnotesize
        \item[5] https://tianchi.aliyun.com/dataset/dataDetail?dataId=42
        \item[6] https://jdata.jd.com/html/detail.html?id=8
         \item[7] https://tianchi.aliyun.com/competition/entrance/231522/information
        \item[8] https://www.kaggle.com/c/acquire-valued-shoppers-challenge/overview
        \item[9] https://tianchi.aliyun.com/dataset/dataDetail?dataId=649
      \end{tablenotes}
\end{threeparttable}
\end{center}
\vspace{-15pt}
\end{table*}

\subsection{Performance of Compared Approaches} 
In order to provide a better reference for fair comparison, Table \ref{table: statisticsofallmehtods} shows the performance of eight approaches across seven metrics on the three datasets \footnote{The approaches in the table are sorted by their publication year.}.

\textit{\textbf{Finding 1:} Although conventional approaches are simple, they can achieve wonderful performance.}
Among the conventional approaches including Top, FPMC, UP-CF@r and TIFUKNN, the two former ones perform similarly on all datasets. As the simplest approach, TOP even performs better than FPMC w.r.t Recall, MAP, MRR and NDCG in most cases. This result confirms the importance of item frequency information in next-basket recommendation. Besides, this demonstrates that TOP should always serve as a baseline when a new model is proposed \cite{ariannezhad2022recanet}. On Instacart and Dunnhumby, UP-CF@r and TIFUKNN are the best performing approaches on all metrics. The two approaches are similar in that they both combine the attributes of items (i.e., recency and frequency) with a user-based nearest neighbor idea. This implies that though the conventional approaches are simple, they are also able to show powerful ability on suitable datasets. 

\textit{\textbf{Finding 2:} The deep neural network based approaches fail to demonstrate consistent and stable performance.} Among the deep neural network based approaches including DREAM, Beacon, Sets2Sets and CLEA, CLEA can automatically capture interactions between historical items and the target item, which achieves the best performance on TaFeng, but it is not the best one for any of the other two datasets. We suspect that it is related to the characteristics of datasets because the average basket size (i.e., items/basket) of TaFeng is lower than those of Instacart and Dunnhumby (see from Table \ref{tab:statistics_datasets}). Sets2Sets utilizes repeated pattern based on encoder-decoder architecture, which outperforms DREAM, Beacon and even CLEA in Dunnhumby. This indicates that there exist numerous repeated purchases in NBR task.

\textit{\textbf{Finding 3:} The conventional approaches can beat the deep neural network based ones on most datasets.}
Regarding the different compared approaches, we can observe that their performances change a lot across datasets, and there is no approach outperforms all others on each dataset. So we further analyze the performance on three datasets in terms of two key metrics, i.e., F1-Score and NDCG, as shown in Figure \ref{fig:performaneF1Ndcg}. We find that Sets2Sets, UP-CF@r, TIFUKNN and CLEA stand out among other approaches. Though Sets2Sets and CLEA show better performance than UP-CF@r and TIFUKNN on Tafeng dataset, they are beaten on Instacart and Dunnhumby datasets. Sets2Sets and CLEA are deep neural based approaches, while UP-CF@r and TIFUKNN are conventional approaches based on KNN. Considering the basket length of Tafeng is shorter than Instacart and Dunnhumby, this may indicate that Sets2Sets and CLEA are more applicable on the datsets with shorter baskets, however, UP-CF@r and TIFUKNN can effectively recommend next basket on the datasets with longer baskets. 

According to the experimental data, we find that under same experimental settings and on the same datasets, the performances of deep neural approaches are inferior to those of conventional ones in most cases.




\section{NBR Algorithms and Datasets}
For facilitating the access for empirical analysis, in table \ref{table: statisticsofallalgorithms}, we summarize source codes of algorithms for NBRs which utilize different models. The listed NBR algorithms are publicly accessible and commonly used as baselines in existing work. 
In addition to algorithms, datasets are necessary for evaluating NBRs algorithms. In order to facilitate further analysis of the investigated algorithms, in table \ref{table: statisticsofalldatasets}, we also summarize datasets for NBRs which can be built as baskets in ascending order in terms of users’ shopping visits.

\section{Conclusion}
In this paper, we have systematically investigated and evaluated the representative works on NBR task. Specifically, we clarify the difference between NBRs and traditional RSs, and formulate the problem statement of NBR. Then, we categorize the existing work into three groups, i.e., conventional approaches, latent representation based approaches and deep neural network based approaches, whose advantages and disadvantages are analyzed. 
With the aim of better understanding and systematical evaluation on NBRs,
we further analyze the popular NBR algorithms and evaluation metrics, and re-run them on the same dataset with unified experimental settings to compare their performance. We have provided a unified framework to fairly evaluate different NBR approaches, which can be utilized as a valuable reference for the related research on NBR. We hope that this work can make readers understand NBR more clearly and make the evaluation of NBRs more fairly and effectively. 

\section*{Acknowledgment}
The work is partly supported by National Nature Science Foundation of China (61502259), and Key Program of Science and Technology of Shandong (2020CXGC010901), and Studio Project of the Research Leader in Jinan (2019GXRC062).

\bibliographystyle{IEEEtran}
\bibliography{ref}

\vspace{12pt}
\color{red}

\end{CJK}
\end{document}